# Atomic-Scale Phonon Engineering for Enhanced Thermal Transport in Wafer-Scale-Bonded β-$Ga_2O_3$/SiC Heterostructures

*Jiaxin Liu, Ruilin Mao, Zhetong Liu, Yuan Yin, Sheng Li, Fangyuan Sun, Bingyao Liu,* Zhenzhong Wang,* and Peng Gao**

J. Liu, R. Mao, Z. Liu, Y. Yin, B. Liu, P. Gao
International Center for Quantum Materials, and Electron Microscopy Laboratory, School of Physics, Peking University, Beijing 100871, China
J. Liu, S. Li, P. Gao
Academy for Advanced Interdisciplinary Studies, Peking University, Beijing 100871, China
F. Sun
School of Energy and Environmental Engineering, University of Science and Technology Beijing, Beijing 100083, China
Z. Wang
Suzhou Ultra-Wafer Co., Ltd., Suzhou, Jiangsu 215100, China
P. Gao
Interdisciplinary Institute of Light-Element Quantum Materials and Research Center for Light-Element Advanced Materials, Peking University, Beijing 100871, China
P. Gao
Tsientang Institute for Advanced Study, Zhejiang 310024, China
E-mail: liubingyao@pku.edu.cn; zzwang@ultrawafer.com; pgao@pku.edu.cn

Jiaxin Liu, Ruilin Mao, and Zhetong Liu contributed equally to this work.

Funding: National Natural Science Foundation of China (12604258); New Cornerstone Science Foundation (XPLORER PRIZE).

## Abstract

Bonding semiconductors to substrates with high thermal conductivity is a promising way to solve the heat-dissipation problem of materials that transport heat poorly, and it avoids the lattice-matching constraints of conventional epitaxy. However, the lattice mismatch remains, leaving the two materials with mismatched phonon states, while poor interface quality scatters phonons even more. Together, these effects block phonon transport across the bonded interface and make interfacial thermal resistance a critical bottleneck, calling for careful interface engineering. Here, β-$Ga_2O_3$/SiC serves as a model system for a phonon-bridging strategy. A nanoscale crystalline Si interlayer is designed and fabricated by wafer bonding to bridge the mismatched phonon states of β-$Ga_2O_3$ and SiC. Spatially resolved electron energy-loss spectroscopy and molecular dynamics simulations show that this Si layer opens new channels for phonon transport across the interface. The engineered interface raises the interfacial thermal conductance to 165 MW $m^{-2}$ $K^{-1}$—the highest value yet reported for β-$Ga_2O_3$/SiC interfaces. These results offer an effective way to manage heat in β-$Ga_2O_3$ devices and point to a general approach for building phonon-bridging interfaces that improve heat dissipation in heterogeneous systems.

## 1. Introduction

β-$Ga_2O_3$ has an ultra-wide bandgap of 4.8 eV and a high breakdown field of about 8 MV $cm^{-1}$, making it a strong candidate for next-generation power and radio-frequency electronics.[1–5] However, its low thermal conductivity of only 10-27 W $m^{-1}$ $K^{-1}$ limits its use at high power: poor heat spreading causes strong self-heating, which degrades both device performance and long-term reliability.[6,7] A promising solution is to bond β-$Ga_2O_3$ to a substrate that conducts heat well. For this purpose, SiC—with a thermal conductivity of roughly 350-490 W $m^{-1}$ $K^{-1}$—is an attractive heat-spreading substrate.[8–10] Conventional heteroepitaxy requires the semiconductor and the substrate to have matching lattices and thermal expansion, which is hard to achieve for materials as different as β-$Ga_2O_3$ and SiC. Their large mismatch creates threading dislocations, twins, and other defects that lower crystal quality and thus degrade device performance.[11,12] Wafer bonding avoids these problems:[13–15] it allows high-quality single-crystal β-$Ga_2O_3$ to be bonded directly onto the substrate without damaging the crystal.[16,17]

Even though wafer bonding relaxes the need for lattice matching, the lattice mismatch itself remains, giving the two materials mismatched phonon states that block phonon transport across the interface and make interfacial thermal resistance a new bottleneck.[18–21] Much effort has gone into engineering interfacial phonon modes with bonding interlayers, including $SiN_x$, $Al_2O_3$, and $SiO_x$, progressively improving the interfacial thermal conductance (ITC) from 21.2 to 162.1 MW $m^{-2}$ $K^{-1}$.[22–25] However, thick interlayers add thermal resistance of their own, while structural disorder can scatter phonons and impede heat transport. Further increases in ITC therefore require careful control of interlayer composition, structure and thickness. Theoretical analyses and atomistic simulations show that suitable interlayers can improve phonon spectral matching and enhance interfacial heat transport.[26,27] These findings motivate thin interlayers with well-matched phonon modes that can bridge the phonon states of the two materials and boost heat flow across the interface.

Here, guided by molecular dynamics (MD) simulations, we bond β-$Ga_2O_3$ to SiC at the wafer scale using a modified surface-activated bonding (SAB) process with an ultrathin (<5 nm) Si interlayer. Using scanning transmission electron microscopy (STEM) with electron energy-loss spectroscopy (EELS), we find that the presence of the very thin Si interlayer enhances the overlap of the phonon density of states (PDOS) across the interface, bridging the mismatched phonons of β-$Ga_2O_3$ and SiC and providing an efficient pathway for heat transport. Time-domain thermoreflectance (TDTR) measurements give an ITC of ~165 MW $m^{-2}$ $K^{-1}$—the highest value reported for β-$Ga_2O_3$/SiC interfaces, which is a large improvement compared to the sample without the Si layer ~86 MW $m^{-2}$ $K^{-1}$. With such an interface, finite-element simulations predict that the channel hotspot temperature of β-$Ga_2O_3$ metal-oxide-semiconductor field-effect transistors (MOSFETs) drops by about 50 K at 4 W $mm^{-1}$ relative to the device without the Si interlayer. This work offers an effective way to manage heat in β-$Ga_2O_3$ devices. It also provides a general strategy for engineering interfacial thermal transport between dissimilar materials, even when their lattices, thermal properties, and phonon states all differ.

## 2. Results and Discussion

### 2.1. Theoretical Design and Wafer-Scale Realization of Si-Assisted β-$Ga_2O_3$/SiC Interfaces

We first investigated the effect of an ultrathin Si interlayer on vibrational coupling between β-$Ga_2O_3$ and SiC. Figure 1a illustrates the heat-transport models designed for the directly bonded and Si-assisted interfaces. At the directly bonded interface, the pronounced vibrational mismatch between β-$Ga_2O_3$ and SiC is expected to cause substantial phonon reflection, thereby restricting heat dissipation into the SiC substrate. Si-interlayer insertion introduces intermediate vibrational states that can couple sequentially with both materials, providing a potential pathway for cross-interface phonon transmission. To evaluate this design, the normalized phonon densities of states of β-$Ga_2O_3$, SiC and Si were calculated (Figure 1b).

$\beta$-$Ga_2O_3$ and SiC exhibited distinct spectral distributions, whereas Si contributed additional vibrational states at low and intermediate energies. Below approximately 60 meV, Si showed greater spectral overlap with $\beta$-$Ga_2O_3$ while retaining overlap with the low-energy modes of SiC. This intermediate spectral distribution suggests that the Si interlayer can reduce the vibrational discontinuity between the two bonded materials. Cumulative PDOS-overlap analysis further quantified the spectral matching among the three materials (Figure 1c). Below 60 meV, the overlap factors of Si with $\beta$-$Ga_2O_3$ and SiC reached 41.5% and 33.0%, respectively, both exceeding that of the direct $\beta$-$Ga_2O_3$/SiC interface, which predicts a continuous phonon-coupling pathway through the Si interlayer.

Guided by the theoretical insights from the calculation, a wafer-scale $\beta$-$Ga_2O_3$/SiC heterostructure was fabricated through a modified Si-assisted SAB process. Before bonding, the $\beta$-$Ga_2O_3$ and SiC wafers were polished to sub-nanometer surface roughness (<0.5 nm). This method introduces an ultrathin Si layer onto the activated SiC surface before contact with the activated $\beta$-$Ga_2O_3$ wafer. The process produced a 2-inch $\beta$-$Ga_2O_3$ wafer bonded onto a 6-inch SiC substrate, with a bonded area exceeding 95% and only small unbonded regions near the wafer edge (Figure 1d). Cross-sectional scanning electron microscopy (SEM) showed a flat and continuous $\beta$-$Ga_2O_3$/SiC interface without visible cracks, delamination or microscale voids in the observed region (Figure 1e). Atomic-resolution high-angle annular dark-field (HAADF) images further revealed ordered $\beta$-$Ga_2O_3$ and SiC lattices along the [010] and $[11\bar{2}0]$ zone axes, respectively, separated by a continuous interfacial layer of approximately 4.4 nm (Figure 1f).

Post-bond annealing at 450 °C for 6 h was then applied to regulate interfacial bonding and structural order. Four representative samples were prepared to decouple the effects of Si insertion and annealing: pristine bonded GaOSiC-P, annealed GaOSiC-A, Si-assisted bonded GaOSiC-Si and Si-assisted bonded plus annealed GaOSiC-Si-A. X-ray diffraction (XRD) $2\theta$-$\omega$ scans showed preserved $\beta$-$Ga_2O_3$ reflections without discernible peak shifts after annealing, indicating no detectable phase change or macroscopic lattice distortion. The higher-order reflections became more pronounced, while the rocking-curve width of the $\beta$-$Ga_2O_3$ (400) reflection decreased by 25%, indicating improved crystalline ordering (Figure S1a). Williamson-Hall analysis further yielded an apparent out-of-plane microstrain approximately 5.4-fold lower in the bonded $\beta$-$Ga_2O_3$ single crystal than in the reference epitaxial film (Figure S1b-d). Tensile testing consistently resulted in cohesive cleavage within the $\beta$-$Ga_2O_3$ bulk, rather than delamination at the $\beta$-$Ga_2O_3$/SiC interface (Figure S2). Because the bonded interface remained intact, the measured fracture stress of approximately 5-6 MPa represents only a lower bound for the interfacial bond strength. These results demonstrate that Si-assisted SAB provides a large-scale, robustly bonded platform with preserved crystalline quality, enabling subsequent investigation of interfacial reconstruction and thermal transport.

### 2.2. Si-Interlayer-Mediated Reconstruction of Bonded Interfaces

Following the interface design and wafer-scale bonding established in Figure 1, we examined how Si-interlayer insertion alters annealing-induced interfacial reconstruction. Both bonding routes included an annealed condition and an unannealed reference: GaOSiC-A and GaOSiC-Si-A were annealed at 450 °C for 6 h, whereas GaOSiC-P and GaOSiC-Si were retained without annealing (Figure S3). Cross-sectional high-resolution transmission electron microscopy (HRTEM) revealed a clear structural difference between the two annealed interfaces. GaOSiC-Si-A contained a continuous interfacial layer of approximately 4.2 nm, including a locally crystallized Si region of approximately 2.9 nm adjacent to SiC (Figure 2a). Resolved lattice fringes and discrete spots in the corresponding fast Fourier transform (FFT) supported this assignment. GaOSiC-A instead retained an approximately 3.1-nm amorphous interlayer, which produced a diffuse FFT ring (Figure 2b). Before annealing, GaOSiC-Si exhibited an approximately 4.7-nm disordered interfacial layer, whereas GaOSiC-P contained locally unbonded regions (Figure S3). These controls show that annealing reorganized both interfaces, but local Si crystallization was observed only in the sample containing the Si interlayer. The compositional profiles followed the same structural contrast. Energy-dispersive X-ray spectroscopy (EDS) identified a distinct Si-rich region near the SiC side of GaOSiC-Si-A, whereas GaOSiC-A showed smoother elemental transitions without additional Si enrichment (Figure 2c,d). Low-loss EELS also distinguished the two interfacial states. GaOSiC-A exhibited a $SiO_x$-related response, while GaOSiC-Si-A showed a localized feature near 17 eV, consistent with the bulk plasmon response of crystalline Si (Figure 2e). The microscopy and spectroscopic evidence therefore support the formation of a locally crystallized Si layer after annealing at 450 °C. This relatively low crystallization temperature may reflect nanoscale confinement and templating by the adjacent SiC lattice.[28,29]

The directly bonded samples followed a different reconstruction pathway. Core-loss O K-edge EELS showed that GaOSiC-A retained a higher oxygen signal across the transition region than GaOSiC-P before the signal decreased rapidly toward the SiC bulk (Figure S4). This profile is consistent with annealing-induced oxygen redistribution near the interface. The accompanying lattice response was evaluated from the interface-normal strain component, $\varepsilon_{yy}$, derived from changes in the y-direction spacing between adjacent β-$Ga_2O_3$ atomic planes. Each point in Figure 2f represents the median value for one atomic plane, while the shaded region spans the corresponding 25%-75% range. GaOSiC-A developed a pronounced positive $\varepsilon_{yy}$ near the interface, indicating local tensile distortion. No comparable increase was observed in GaOSiC-Si-A. The corresponding high-angle annular dark-field scanning transmission electron microscopy (HAADF-STEM) images and geometric phase analysis (GPA) maps are shown in Figure S5. Previous studies have reported that annealing-induced interdiffusion and oxygen-related defects can perturb local atomic coordination and produce lattice distortion in β-$Ga_2O_3$-based interfaces.[23,30,31] The concurrent oxygen redistribution and positive $\varepsilon_{yy}$ suggest that oxygen diffusion may contribute to the tensile distortion in GaOSiC-A.

The two interfaces therefore evolved along different pathways. The directly bonded interface remained predominantly amorphous and underwent oxygen redistribution accompanied by local tensile distortion. Si-interlayer insertion instead enabled local Si crystallization and reduced the lattice distortion in the adjacent β-$Ga_2O_3$. These structural differences provide the basis for the subsequent analysis of interfacial phonon transport.

### 2.3. Phonon Bridging through a Locally Crystallized Si Interlayer

The distinct interface structures identified in Figure 2 prompted us to examine how Si-interlayer insertion affects local vibrational coupling. Spatially resolved EELS was used to map phonon excitations across the two annealed interfaces (Figure 3a-d). At the GaOSiC-A interface, the low-energy β-$Ga_2O_3$ mode shifted toward lower energy (Figure 3a,b). Based on the established sensitivity of phonon energy to lattice strain, this softening is attributed mainly to the tensile lattice expansion previously identified near the interface.[32] By contrast, GaOSiC-Si-A exhibited distinct interfacial modes near 28 and 55 meV, together with a shift of the adjacent SiC phonon branches toward lower energies near the interface (Figure 3c,d). The unannealed controls showed a broader transition region in GaOSiC-P and less defined Si-related features in GaOSiC-Si (Figure S6), indicating that annealing sharpened the local vibrational response of the Si interfacial layer. To quantify vibrational matching, we calculated spectral overlaps from the area-normalized experimental phonon spectra (Figure 3b,d). From GaOSiC-A to GaOSiC-Si-A, the interface/β-$Ga_2O_3$ overlap increased from 62.6% to 68.6%, and the interface/SiC overlap increased from 70.1% to 73.2%. The overlap shared by all three spectra increased from 41.7% to 56.0%. These results show that the locally crystallized Si interlayer provides stronger spectral overlap between β-$Ga_2O_3$ and SiC.

MD simulations were then used to identify the origin of these interfacial modes. The PDOS calculated for GaOSiC-Si-A showed that the Si phonon bridge introduced intermediate vibrational states overlapping with both β-$Ga_2O_3$ and SiC (Figure 3e). Spatially resolved phonon spectra calculated from the annealed atomistic models also reproduced localized bridging features in GaOSiC-Si-A that were absent from GaOSiC-A (Figure S7a,b). These results support the assignment of the experimentally observed interface features to vibrational modes associated with the locally crystallized Si interlayer. The atomistic interface models used for the MD phonon and heat-transport calculations are shown in the inset of Figure 3f. In particular, the GaOSiC-Si-A model explicitly incorporated a locally crystallized Si interlayer to represent the interface structure observed experimentally. Stable heat flows were then established in the two annealed models, as indicated by the nearly linear cumulative energy changes in the heat-source and heat-sink regions (Figure S7c,d). On this basis, the spectrally decomposed ITC was calculated for GaOSiC-A and GaOSiC-Si-A (Figure 3f). For both interfaces, the main thermal contribution occurred below approximately 60 meV. Two additional contributions within 20-40 and 40-60 meV appeared in GaOSiC-Si-A but were absent from GaOSiC-A. Together with the PDOS and EELS results, these contributions were attributed to Si-related vibrational modes that provide additional channels for cross-interface heat transport. Below 20 meV, a transport channel was present in both models but was markedly enhanced in GaOSiC-Si-A. The Si phonon bridge therefore not only contributed intermediate-energy modes but also strongly activated the existing low-energy transport channel. These combined contributions account for the improved phonon transmission across the locally crystallized Si interlayer.

### 2.4. High Interfacial Thermal Conductance and Device-Level Thermal Benefits

We next quantified their macroscopic thermal effect using TDTR. In this pump-probe measurement, a metal transducer is optically heated and its time-dependent thermoreflectance response is fitted using a multilayer heat-diffusion model to extract the thermal boundary resistance (TBR) and ITC of the buried interface.[33–35] The directly bonded and annealed GaOSiC-A interface exhibited an ITC of 86 MW $m^{-2}$ $K^{-1}$ (Figure 4a). For comparison, an ITC of approximately 105 MW $m^{-2}$ $K^{-1}$ has been reported for a directly bonded β-$Ga_2O_3$/SiC interface after annealing.[24] With the Si interlayer, the ITC of GaOSiC-Si-A increased to 165

MW $m^{-2}$ $K^{-1}$, representing an enhancement of 91.8% over GaOSiC-A. This value is among the highest reported for β-$Ga_2O_3$/SiC interfaces in peer-reviewed studies, comparable to the 162.1 MW $m^{-2}$ $K^{-1}$ achieved in an atomic-layer-deposited heterostructure with an anneal-induced covalent interface.[25] It also exceeds the 133 MW $m^{-2}$ $K^{-1}$ reported for SAB-bonded interfaces with an $Al_2O_3$ interlayer.[24] Figure 4b shows a representative TDTR measurement curve for GaOSiC-Si-A. The best fit yielded a TBR of approximately 6.06 $m^2$ K $GW^{-1}$, corresponding to an ITC of 165 MW $m^{-2}$ $K^{-1}$. Changing the TBR by ±20% from the best-fit value produced discernible deviations from the measured response, confirming that the TDTR signal was sensitive to the buried-interface resistance within the selected fitting window.

To evaluate the device-level thermal effect of the enhanced ITC, the measured values were incorporated into a COMSOL Multiphysics finite-element model of a β-$Ga_2O_3$/SiC MOSFET. The device geometry, material properties, heat-loading conditions and boundary conditions were held constant, with only the interface conductance changed from 86 to 165 MW $m^{-2}$ $K^{-1}$. At a dissipated power of 4 W $mm^{-1}$ shown in Figure 4c, the higher-ITC interface promoted heat transfer from the β-$Ga_2O_3$ active layer into the SiC substrate and reduced the simulated peak temperature by approximately 50 K (from 469 to 421 K). The power-dependent simulations in Figure 4d further distinguish the effects of the substrate and interfacial thermal conductance. At the same dissipated power, both GaOSiC-A and GaOSiC-Si-A exhibit substantially lower peak temperature rises above room temperature than the β-$Ga_2O_3$/β-$Ga_2O_3$ reference structure, demonstrating the thermal benefit of the SiC substrate. GaOSiC-Si-A consistently shows a lower temperature rise than GaOSiC-A, indicating that enhanced interfacial heat transport provides an additional improvement in heat dissipation. The model geometry and boundary conditions are described in the Experimental Section.

Taken together, the TDTR measurements and device simulations establish the locally crystallized Si interlayer as an effective strategy for reducing the TBR of bonded β-$Ga_2O_3$/SiC interfaces. By providing intermediate vibrational states while maintaining a structurally connected interface, the Si interlayer offers a practical design principle for improving heat dissipation in β-$Ga_2O_3$ devices. However, the effect of the Si interlayer thickness has not yet been systematically resolved, partly because precise control of an ultrathin buried layer during bonding remains experimentally challenging. Further improvement may depend on balancing the transport resistance associated with the Si layer, including its thickness relative to the effective mean free paths of the heat-carrying phonons, against the interfacial continuity and local crystallinity required to sustain phonon bridging. Systematic thickness-dependent measurements and atomistic calculations will therefore be needed to identify the optimum interlayer configuration.

## 3. Conclusion

We demonstrated a simulation-guided Si-assisted bonding strategy for enhancing thermal transport across wafer-scale-bonded β-$Ga_2O_3$/SiC interfaces. PDOS calculations first identified Si as a suitable phonon-bridging layer, and a modified surface-activated bonding process enabled its wafer-scale integration as an ultrathin interlayer, followed by annealing at 450 °C for 6 hours to induce local crystallization. Spatially resolved EELS revealed that the Si phonon modes near 28 and 55 meV bridge the phonon gap between β-$Ga_2O_3$ and SiC, increasing the phonon spectral-overlap coefficient from 41.7% to 56.0%. TDTR measurements yielded an ITC of 165 MW $m^{-2}$ $K^{-1}$ for GaOSiC-Si-A (with the Si interlayer), which is much higher than that of GaOSiC-A (without the Si interlayer) and is the highest value reported for β-$Ga_2O_3$/SiC interfaces. Incorporating the measured conductance into finite-element device simulations reduced the predicted hotspot temperature by ~50 K within the operating-temperature range of β-$Ga_2O_3$ electronics. More broadly, this work establishes the design and engineering of phonon-bridging interlayers as a general strategy for overcoming interfacial thermal-transport bottlenecks in lattice-mismatched bonded heterostructures, providing a pathway toward efficient thermal management in heterogeneous electronics.

## 4. Experimental Section

### 4.1. Sample Preparation

The 6-inch 4H-SiC and 2-inch β-$Ga_2O_3$ wafers have crystallographic orientations of <0001> and <100>, respectively, each with a tilt angle of less than 0.5°. Before bonding, the mating surfaces were polished to a surface roughness below 0.5 nm. The surfaces were activated by Ar fast atom beam irradiation for 120 s under ultrahigh vacuum (<1 × $10^{-5}$ Pa). For Si-assisted bonding, an ultrathin Si interlayer was deposited on the activated SiC surface before contact; this deposition step was omitted for the directly bonded samples. The wafers were subsequently aligned, brought into contact at room temperature, and pressed under a load of 10000 N for 150 s. The annealed samples were then treated at 450 °C for 6 h under $N_2$. Transmission electron microscopy (TEM) samples were prepared by the focused ion beam (FIB, Helios G4 UX, Thermo Fisher) to observe the cross section.

### 4.2. STEM Characterization and EELS Acquisition

We recorded HRTEM and HAADF images at 300 kV using an aberration-corrected FEI Titan Themis G2. For HAADF-STEM imaging, the probe convergence semi-angle was 30 mrad, and the detector collection semi-angle range was 39-200 mrad. EDS analysis was also acquired by Titan. EELS spectra were recorded using a Nion HERMES 200 microscope at 60 kV. The convergence and collection semi-angles were 35 mrad and 25 mrad. We set the dispersion at 0.4065 eV/ch, 0.032 eV/ch and 0.0005 eV/ch to collect core loss, low loss and phonon signal data, respectively. The total number of acquisition channels is 2048.

### 4.3. MD Calculations

The MD simulations were performed using the graphics processing unit (GPU)-accelerated GPUMD package,[36,37] with interatomic interactions described by the universal neuroevolution potential NEP89.[38] For model construction, amorphous $Ga_2O_3$ and Si slabs were generated from randomly constructed atomic configurations followed by thermal annealing and structural relaxation. A compositionally graded amorphous $Ga_2O_3$/SiC interlayer was also generated from a random atomic configuration with a gradual compositional transition and subsequently relaxed. The resulting amorphous slabs were combined with crystalline 4H-SiC and $Ga_2O_3$, as appropriate, to construct the corresponding interface models. Prior to the production simulations, each constructed interface model was further relaxed using the fast inertial relaxation engine (FIRE) algorithm.[39] The system was then equilibrated at 300 K and zero external pressure in the isothermal-isobaric (NPT) ensemble for 50 ps, followed by a further 50 ps in the canonical (NVT) ensemble, using a time step of 0.5 fs and Berendsen temperature/pressure coupling.[40]

For vibrational analysis, the equilibrated system was propagated in the microcanonical (NVE) ensemble with a time step of 1 fs. After a 100 ps pre-equilibration, atomic velocities were collected for an additional 100 ps. The PDOS was obtained from the Fourier transform of the velocity autocorrelation function,

$$D(\omega) \propto \int_0^{\tau} \langle \mathbf{v}(0) \cdot \mathbf{v}(t) \rangle \, e^{i\omega t} \, dt.$$

The velocity correlation was sampled every 5 fs using 200 correlation points, and the PDOS was evaluated on 500 frequency points up to a maximum angular frequency of 250 THz.

Non-equilibrium molecular dynamics (NEMD) simulations were used to evaluate interfacial thermal transport. Boundary atoms were fixed, and local Langevin thermostats[41] were applied at 500 and 300 K along the transport direction. The system was evolved for 0.5 ns to establish a steady state, followed by a 1 ns production run. The steady-state heat flux and ITC were calculated as

$$J = \frac{\dot{Q}}{A},$$

$$G = \frac{J}{\Delta T_{\mathrm{int}}},$$

where $\dot{Q}$ is the energy-transfer rate, $A$ is the cross-sectional area normal to heat flow, and $\Delta T_{\mathrm{int}}$ is the interfacial temperature discontinuity obtained by linearly extrapolating the temperature profiles on the two sides of the interface.

The frequency-dependent heat transport was further analyzed using the spectral heat current (SHC) formalism implemented in GPUMD.[37,42] The nonequilibrium virial-velocity correlation function and its Fourier transform are

$$\mathbf{K}(t) = \sum_i \langle \mathbf{W}_i(0) \cdot \mathbf{v}_i(t) \rangle,$$

$$\widetilde{\mathbf{K}}(\omega) = \int_{-\infty}^{\infty} \mathbf{K}(t) e^{i\omega t}\, dt,$$

and the spectral heat current is expressed as

$$\mathbf{J}_q(\omega) = 2\widetilde{\mathbf{K}}(\omega).$$

For the SHC calculation, the correlation function was sampled every 2 fs with 250 correlation steps. The transport direction was $z$, and 500 frequency points were used up to a maximum angular frequency of 250 THz.

### 4.4. Device Thermal Modeling and Finite-Element Simulation

To evaluate the heat dissipation efficacy enabled by interfacial engineering, a two-dimensional finite-element thermal model of the β-$Ga_2O_3$ MOSFET was implemented using COMSOL Multiphysics. The device cross-section incorporates an ultra-thin β-$Ga_2O_3$ channel ($t_{GaO}$ = 50 nm) on a host substrate ($H_{sub}$ = 10 μm, $L_{sub}$ = 40 μm), with an active heating region defined by a gate length of $L_G$ = 0.3 μm. Steady-state conduction governed by the Fourier heat equation, $\nabla \cdot (\kappa \nabla T) + Q = 0$, was solved across the domain. To account for concentrated channel Joule heating, a uniform volumetric heat generation rate was applied under the gate:

$$Q = \frac{P_{line}}{L_G \cdot t_{GaO}}$$

where $P_{line}$ represents the linear power density. The β-$Ga_2O_3$/substrate hetero-interface was parameterized via a "Thin Layer" boundary condition (5 nm thickness) using ITC values experimentally measured by TDTR. A homogeneous $Ga_2O_3$/$Ga_2O_3$ structure without interfacial thermal resistance served as the baseline. The substrate base was maintained at 293.15 K, while all other outer boundaries were set as adiabatic. A refined non-uniform mesh (<10 nm element size near the channel and interface) was deployed, with mesh independence confirmed by a peak temperature variation below 0.1%.

### 4.5. TDTR Characterization

TDTR measurements were performed using a Pioneer-ONE system (Autinst Technology Co., Ltd.). The system employs a femtosecond fiber laser with a fundamental wavelength of 1064 nm, a pulse duration of <120 fs and a repetition rate of 80 MHz. The frequency-doubled 532 nm beam was used as the pump, with its intensity modulated at 1 MHz using an electro-optic modulator. Before measurement, a 100 nm Al transducer layer is deposited onto the samples. The pump-probe delay was controlled using a mechanical optical delay line. The reflected probe signal was detected using a high-speed photodetector and demodulated using a lock-in amplifier. The measured signals were fitted using a multilayer thermal-diffusion model to extract the TBR, and the ITC was calculated as ITC = $TBR^{-1}$. To ensure statistical robustness and minimize artifacts arising from surface inhomogeneity or contamination, each sample is measured at three or more distinct locations.

### Supporting Information

Supporting Information accompanies this preprint.

### Acknowledgements

This work was supported by the National Natural Science Foundation of China (12604258 to B.Liu). The authors acknowledge the Electron Microscopy Laboratory of Peking University for electron microscopies. We acknowledge the High-performance Computing Platform of Peking

University for providing computational resources for the calculation. P.G. acknowledges the support from the New Cornerstone Science Foundation through the XPLORER PRIZE.

**Author Contributions**

J.Liu., R.M., and Z.Liu. contributed equally to this work. P.G., Z.W., and J.Liu. conceived the project. Z.Liu. fabricated the bonded samples, and S.Li. performed the post-bond annealing. J.Liu. prepared the TEM sample and performed the EELS and STEM measurements. F.S. performed the TDTR measurements and analyzed the resulting data. R.M. constructed the atomistic interface models and performed the MD simulations, including the PDOS and spectrally decomposed heat-transport analyses. J.Liu. performed the finite-element thermal simulations using COMSOL Multiphysics. R.M., Y. Y., and B.Liu. contributed to the data summary and figure configuration. J.Liu. and P.G. wrote the manuscript. All authors contributed to this work through useful discussion and/or comments to the manuscript. P.G. supervised the project.

**Conflict of Interest**

The authors declare no competing interests.

**Data Availability Statement**

The data that support the plots within this paper and the other findings of this study are available from the corresponding author upon reasonable request.

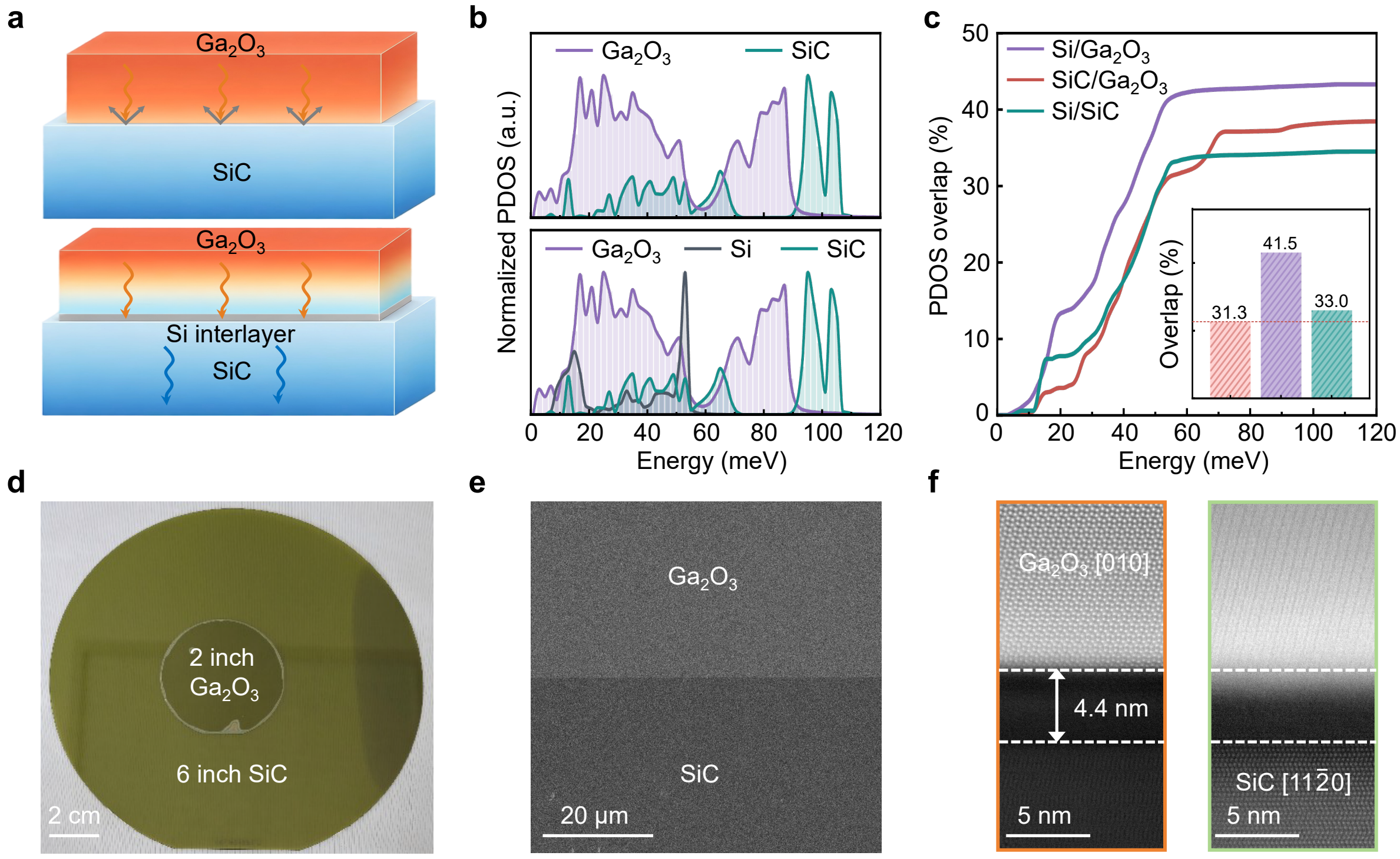

**Figure 1.** Theoretical prediction and structural realization of Si-assisted β-$Ga_2O_3$/SiC interfaces. a) Schematic of phonon reflection at a directly bonded interface and enhanced phonon transmission through an ultrathin Si interlayer. b) Calculated normalized PDOS for $Ga_2O_3$, Si and SiC. c) Cumulative PDOS overlap for different material pairs. The inset shows the overlap below 60 meV. d) Optical photograph of the 2-inch $Ga_2O_3$/6-inch SiC bonded wafer, showing small unavoidable voids at the wafer periphery during bonding. e) Cross-sectional SEM image of the uniform and crack-free $Ga_2O_3$/SiC interface. f) Cross-sectional HAADF image of the Si-assisted bonded interface, viewed along the [010] zone axis of $Ga_2O_3$ (orange box) and the [11$\bar{2}$0] zone axis of SiC (green box), revealing an ~4.4 nm interlayer at the interface.

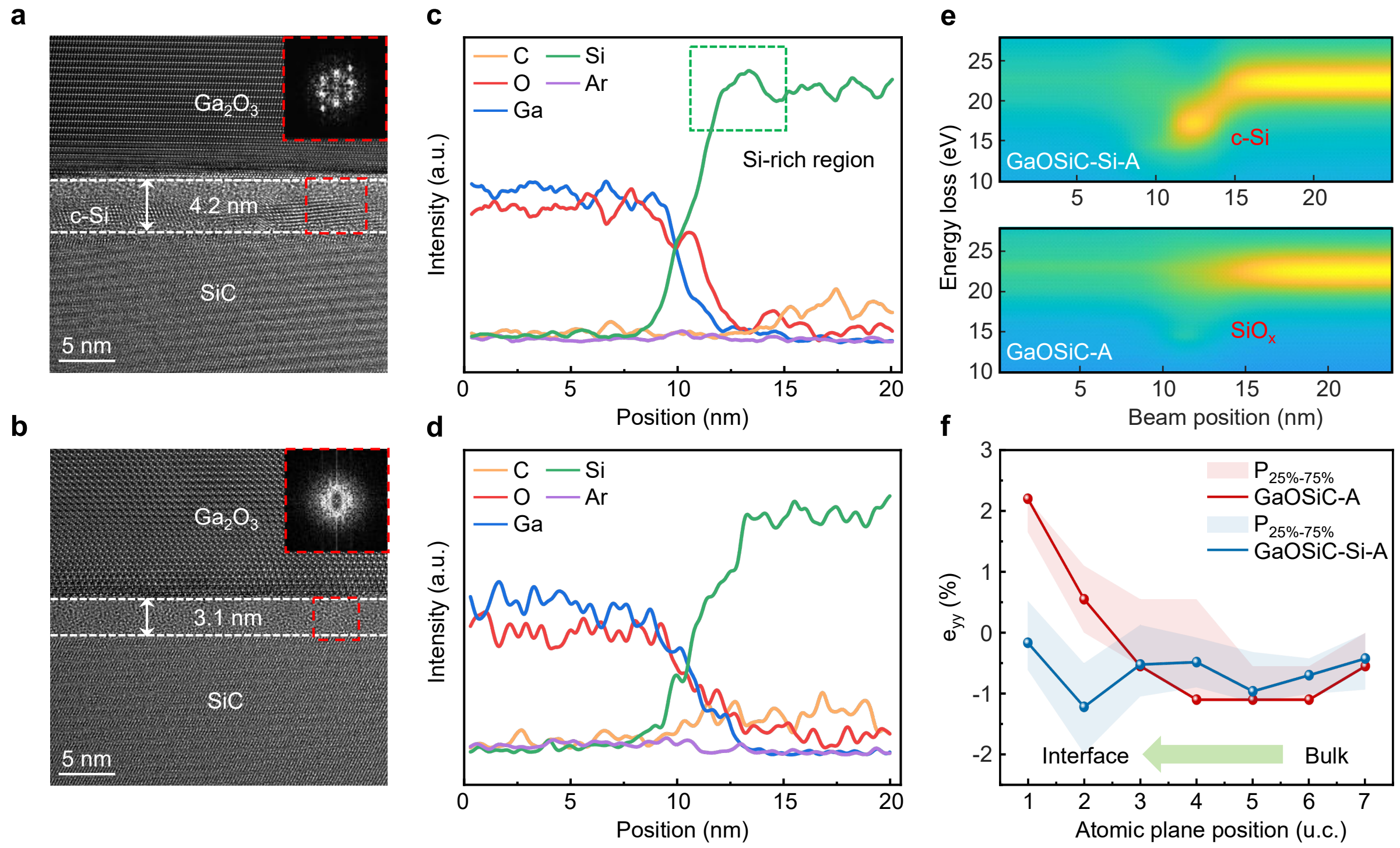

**Figure 2.** Interfacial structural evolution and elemental distribution of $Ga_2O_3$/SiC heterojunctions. a) Cross-sectional HRTEM images of the bonded interfaces in GaOSiC-Si-A and b) GaOSiC-A, with the insets showing the corresponding FFT patterns of the intermediate layers. Distinct recrystallization of the Si interlayer is resolved in a). c), d) Spatial EDS elemental line profiles across the interfaces corresponding to a) and b), respectively. e) Low-loss EELS mappings of GaOSiC-Si-A (top) and GaOSiC-A (bottom). f) Atomic-plane-resolved $\varepsilon_{yy}$ profiles in $Ga_2O_3$ for GaOSiC-A and GaOSiC-Si-A. Lines show the median values, and shaded regions indicate the 25%-75% percentile range.

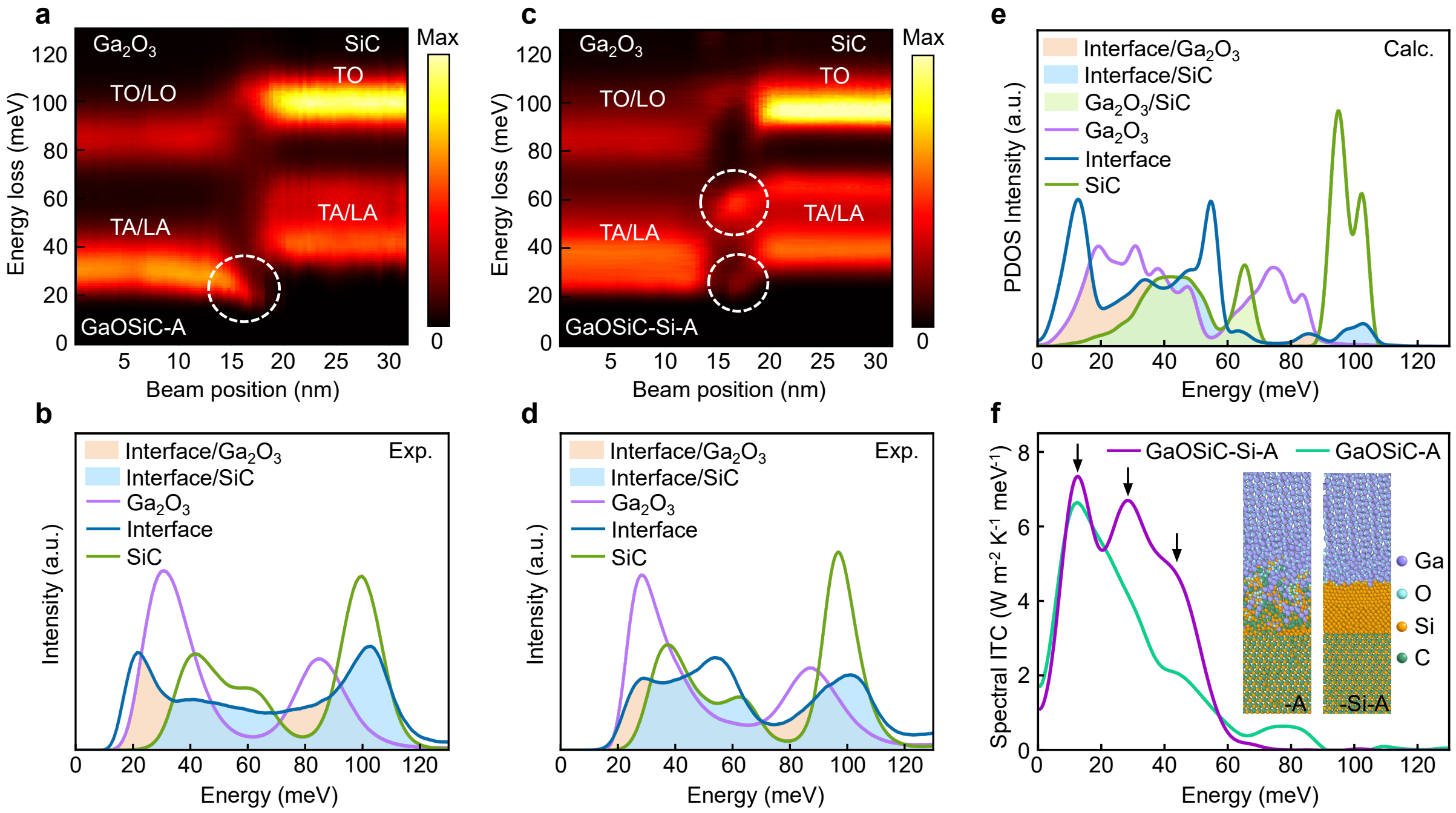

**Figure 3.** Phonon spectrum analysis for interfacial thermal transport. a) Spatially resolved EELS mapping of phonon spectra across the GaOSiC-A interface, and the extracted bulk versus interface phonon spectra b). c), d) Corresponding EELS phonon maps and extracted bulk and interface phonon spectra for the GaOSiC-Si-A interface. The white dashed circles highlight the explicit emergence of a unique localized interfacial vibrational mode. e) MD-simulated vibrational spectra of the GaOSiC-Si-A configuration, with color-shaded regions defining the enhanced cross-interface spectral overlaps. f) Mode-decomposed spectral interfacial thermal conductance for different configurations. Arrows index the direct contribution from the interfacial mode, and the inset shows the corresponding atomistic interface models used in the MD simulations.

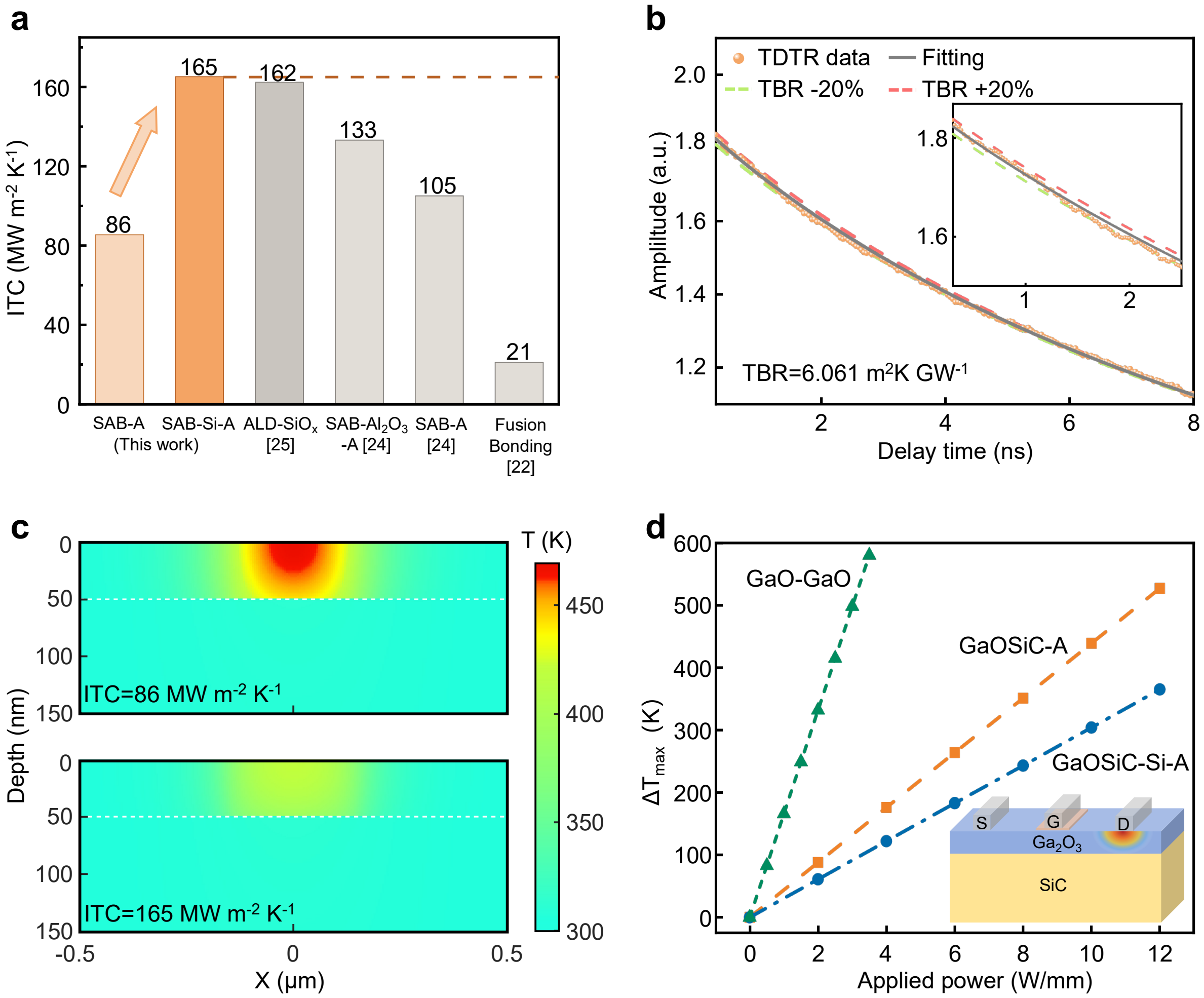

**Figure 4.** TDTR-measured interfacial thermal conductance and finite-element modeling of $Ga_2O_3$/SiC heterointerfaces. a) TDTR-measured ITCs of GaOSiC-A and GaOSiC-Si-A samples compared with reported values for $Ga_2O_3$/SiC interfaces. b) TDTR measurement curve for GaOSiC-Si-A and the corresponding fit, yielding a TBR of 6.061 $m^2$ K $GW^{-1}$. Dashed curves represent ±20% variations in TBR. c) COMSOL-simulated temperature maps under identical operating conditions. d) Simulated peak channel temperature rise relative to room temperature as a function of applied power density for different configurations, and the inset shows the schematic of the β-$Ga_2O_3$ MOSFET structure used for thermal simulation.

# Supporting Information

## Atomic-Scale Phonon Engineering for Enhanced Thermal Transport in Wafer-Scale-Bonded β-$Ga_2O_3$/SiC Heterostructures

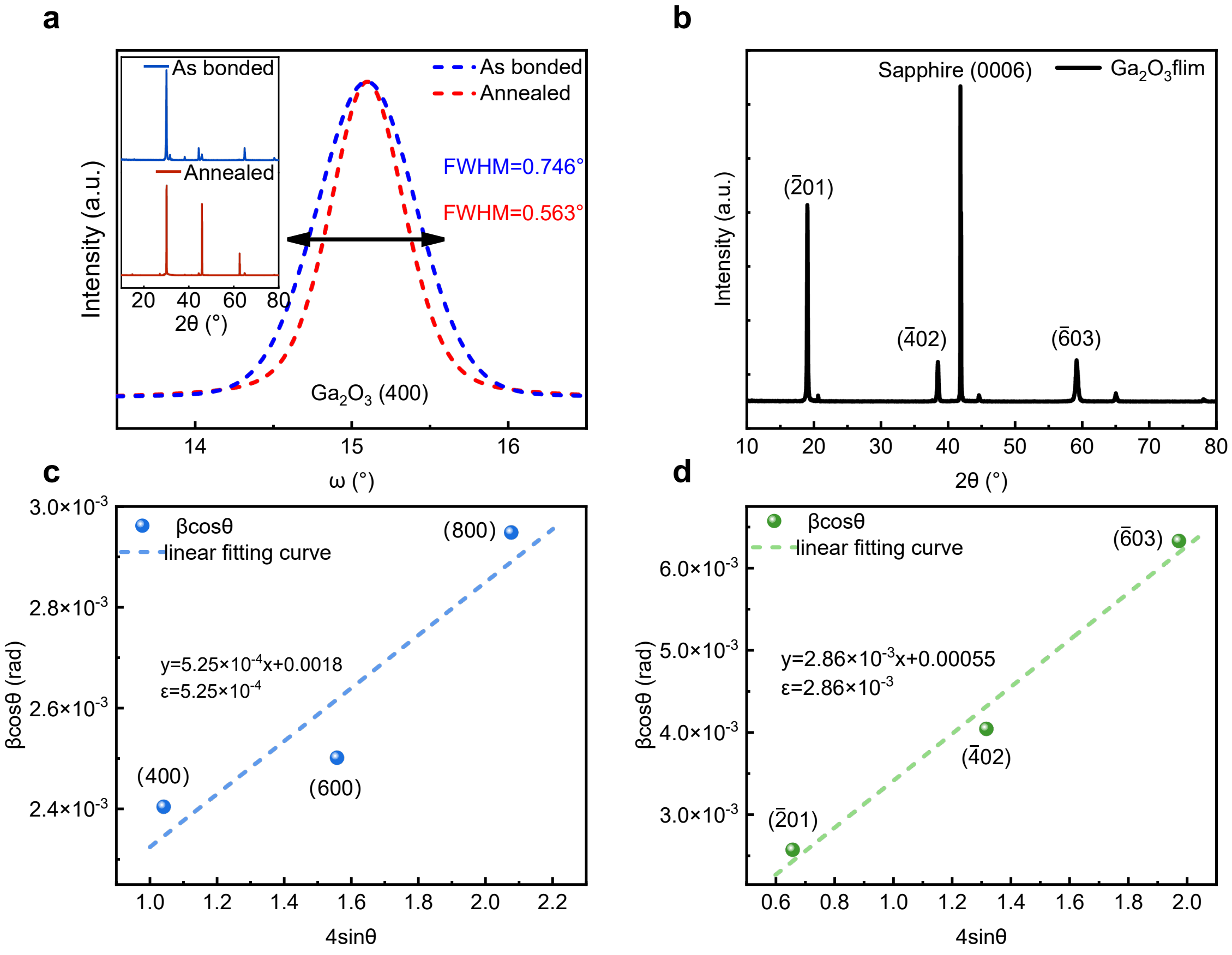

**Figure S1.** Crystalline quality and microstrain of bonded and epitaxial β-$Ga_2O_3$. a) XRD ω-rocking curves of the β-$Ga_2O_3$ (400) reflection before and after post-bonding annealing. The inset shows the corresponding 2θ scans. The reduced rocking-curve width after annealing indicates improved crystalline ordering. b) XRD 2θ scan of the epitaxial β-$Ga_2O_3$ reference film. c), d) Williamson-Hall plots for the bonded β-$Ga_2O_3$ single crystal c) and epitaxial β-$Ga_2O_3$ film d). Linear fits of βcosθ against 4sinθ yield microstrains of $5.25 \times 10^{-4}$ and $2.86 \times 10^{-3}$, respectively, indicating lower microstrain in the bonded single crystal.

The bonded β-$Ga_2O_3$ layer was derived from a (100)-oriented single crystal, whereas the epitaxial reference film exhibited a predominant ($\bar{2}01$) out-of-plane orientation (Figure S1a, b). Because diffraction-peak widths depend on the crystallographic reflection, diffraction geometry, mosaicity and structural anisotropy, the widths of non-equivalent reflections cannot be compared directly as a quantitative measure of crystalline quality. This consideration is particularly important for monoclinic β-$Ga_2O_3$, which exhibits pronounced orientation-dependent X-ray peak broadening.[1] Williamson-Hall analysis was therefore employed to evaluate the peak-broadening contributions and compare the apparent out-of-plane microstrain of the two samples.[2,3]

Under the uniform deformation approximation, the contributions of finite coherent-domain size and lattice microstrain to the diffraction-peak broadening are described by

$$\beta cos\theta = \frac{K\lambda}{D} + 4\varepsilon sin\theta$$

where $\beta$ is the measured diffraction-peak breadth in radians, $\theta$ is the Bragg angle, $K$ is the shape factor, $\lambda$ is the X-ray wavelength, $D$ is the coherent-domain size and $\varepsilon$ is the apparent microstrain.[2,3] Accordingly, linear fitting of $\beta cos\theta$ against $4sin\theta$ yields $\varepsilon$ from the slope, whereas the intercept corresponds to the size-broadening term $K\lambda/D$. Similar analysis has been applied to evaluate strain-related peak broadening in β-$Ga_2O_3$ crystals and thin films.[1,4]

The higher-order (400), (600) and (800) reflections were analyzed for the bonded β-$Ga_2O_3$ single crystal, whereas the ($\bar{2}$01), ($\bar{4}$02) and ($\bar{6}$03) reflections were used for the epitaxial film (Figure S1c,d). The linear fits yielded apparent microstrains of $5.25\times10^{-4}$ and $2.86\times10^{-3}$, respectively. Thus, the bonded single crystal exhibited an approximately 5.4-fold lower apparent microstrain than the epitaxial film. Together with the approximately 25% narrowing of the β-$Ga_2O_3$ (400) rocking curve after annealing, these results indicate reduced lattice-distortion-related broadening and better preserved crystalline quality in the bonded single-crystal layer.

a b

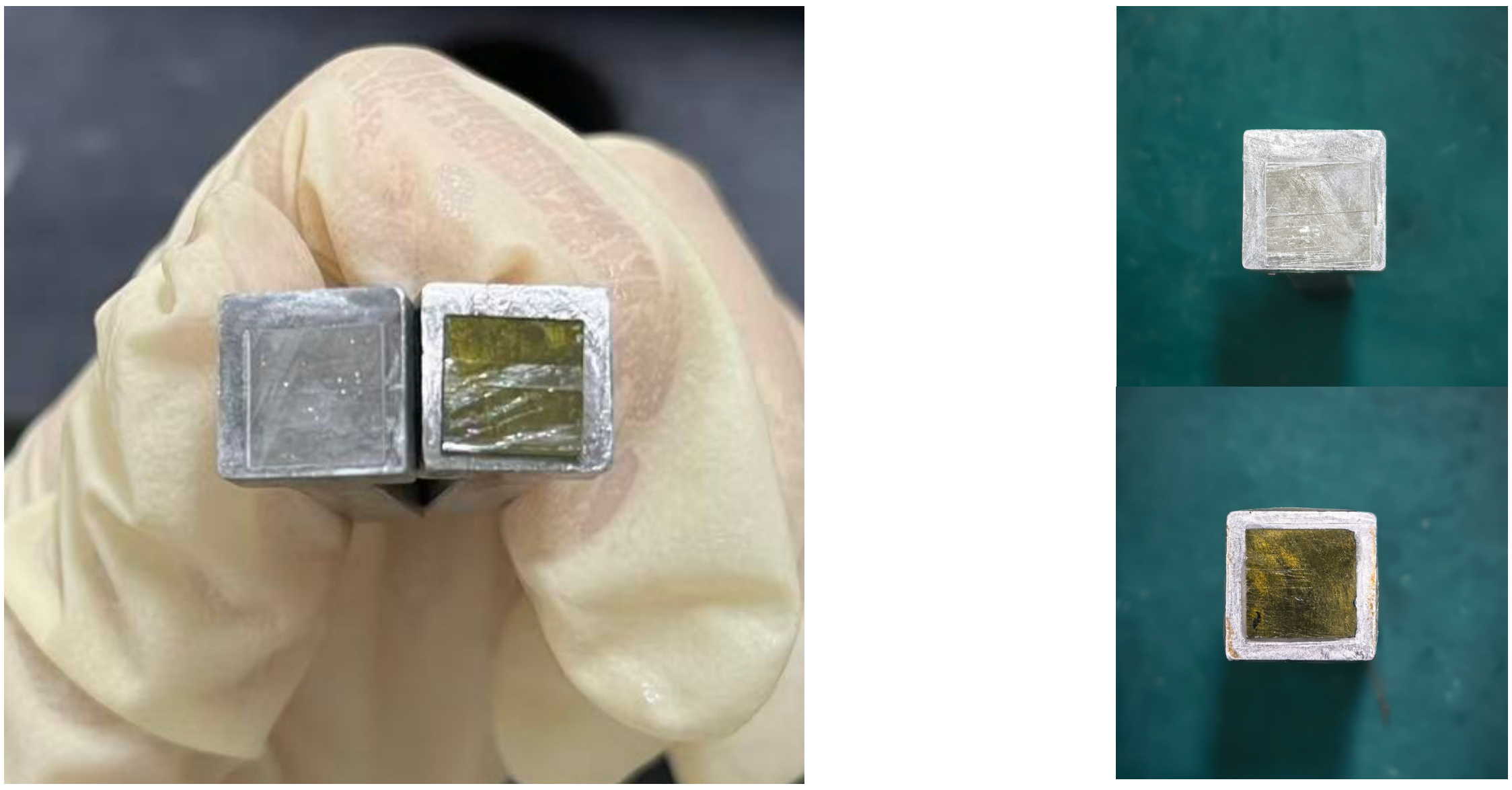

c

| Sample | Tensile Strength（MPa） | Bonding Condition |
|---|---|---|
| **GaOSiC-P** | 5.2 | Interplanar delamination (reference 4-6 MPa) |
| **GaOSiC-A** | 5.6 | |
| **GaOSiC-Si** | 6.3 | |
| **GaOSiC-Si-A** | 5.9 | |

**Figure S2.** Tensile testing of bonded β-$Ga_2O_3$/SiC samples. a), b) Photographs of the separated specimens after tensile testing. Residual β-$Ga_2O_3$ on the SiC-side fracture surfaces is consistent with cohesive cleavage within β-$Ga_2O_3$ rather than failure at the bonded interface. c) Tensile fracture stresses of the four bonded samples. The measured values of 5.2-6.3 MPa are comparable to the reported cleavage strength of β-$Ga_2O_3$ (4-6 MPa),[5–7] indicating that the interfacial bonding strength exceeded the measured cohesive fracture stress.

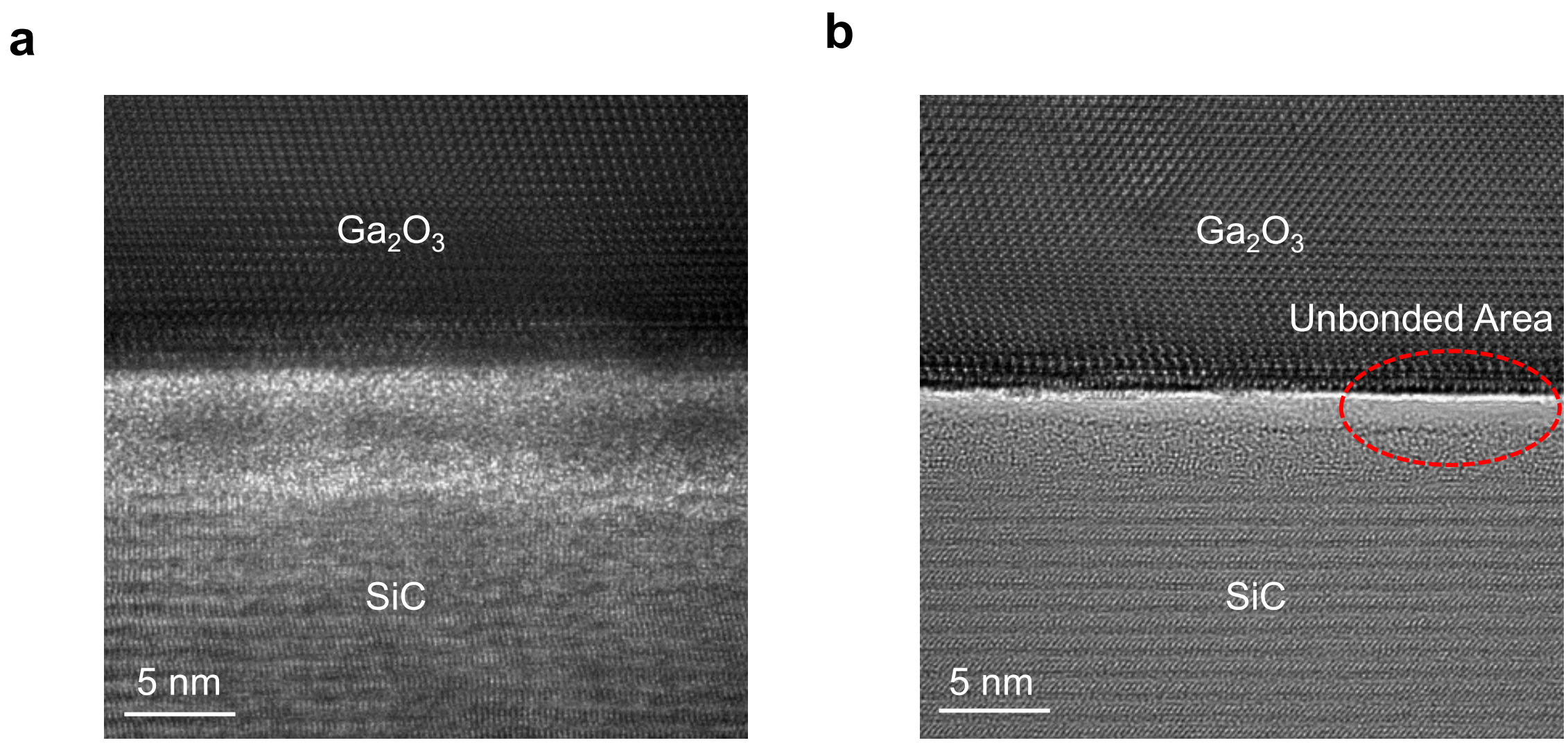

**Figure S3.** Interfacial structures of the unannealed β-$Ga_2O_3$/SiC bonded samples. a), b) Cross-sectional HRTEM images of GaOSiC-Si a) and GaOSiC-P b), showing their interfacial morphologies before annealing. The dashed ellipse in b) indicates a locally unbonded region at the directly bonded interface.

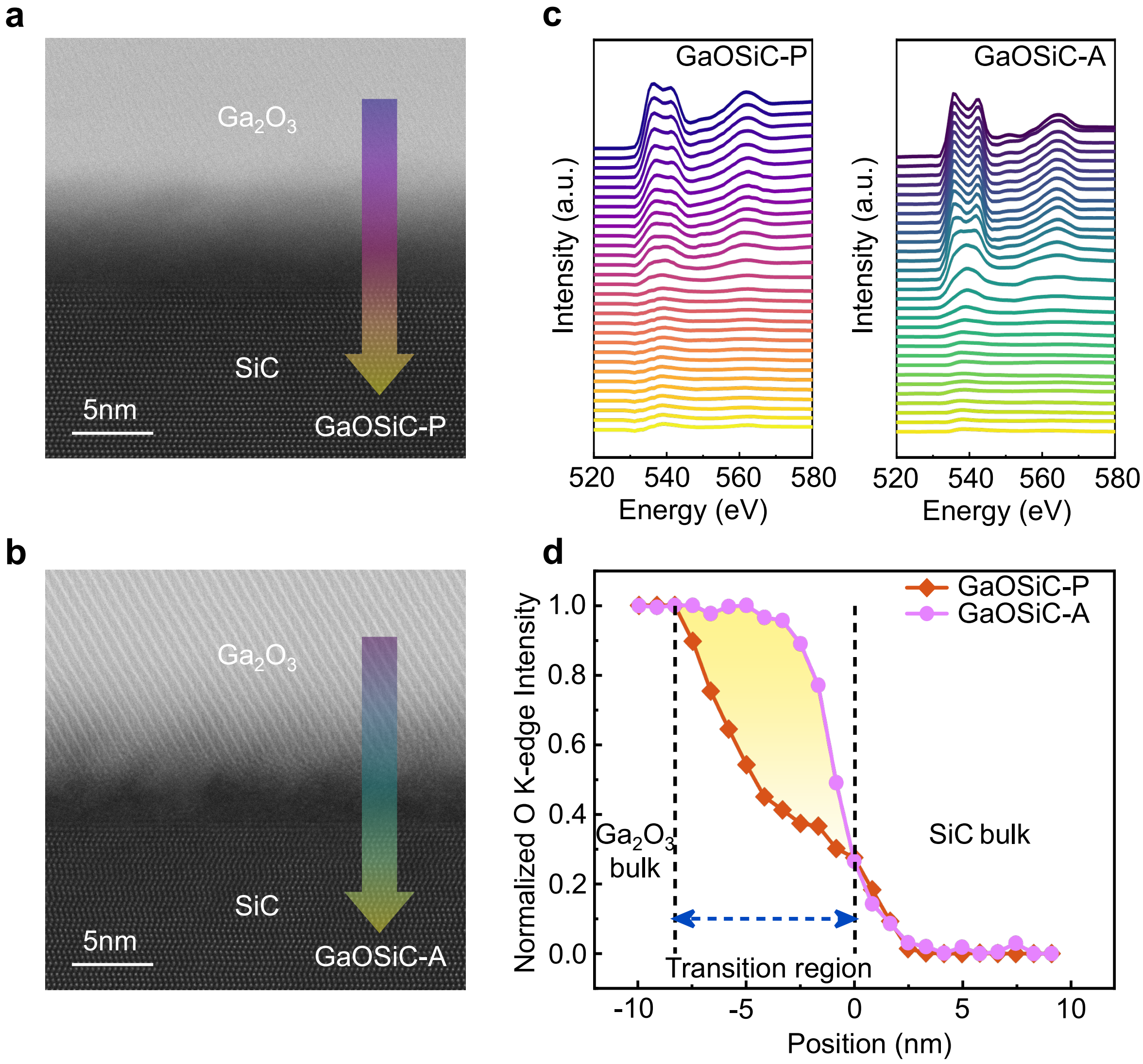

**Figure S4.** Annealing-induced oxygen redistribution at directly bonded β-$Ga_2O_3$/SiC interfaces. a), b) The cross-sectional HAADF image of GaOSiC-P a) and GaOSiC-A b). The colored arrows indicate the O K-edge EELS acquisition direction from β-$Ga_2O_3$ toward SiC. c) Spatially resolved core-loss O K-edge spectra acquired along the arrows in a) and b), with colors corresponding to the probe positions. d) Normalized integrated O K-edge intensity profiles across the interfaces. The increased oxygen signal within the transition region after annealing is consistent with oxygen redistribution toward the interface.

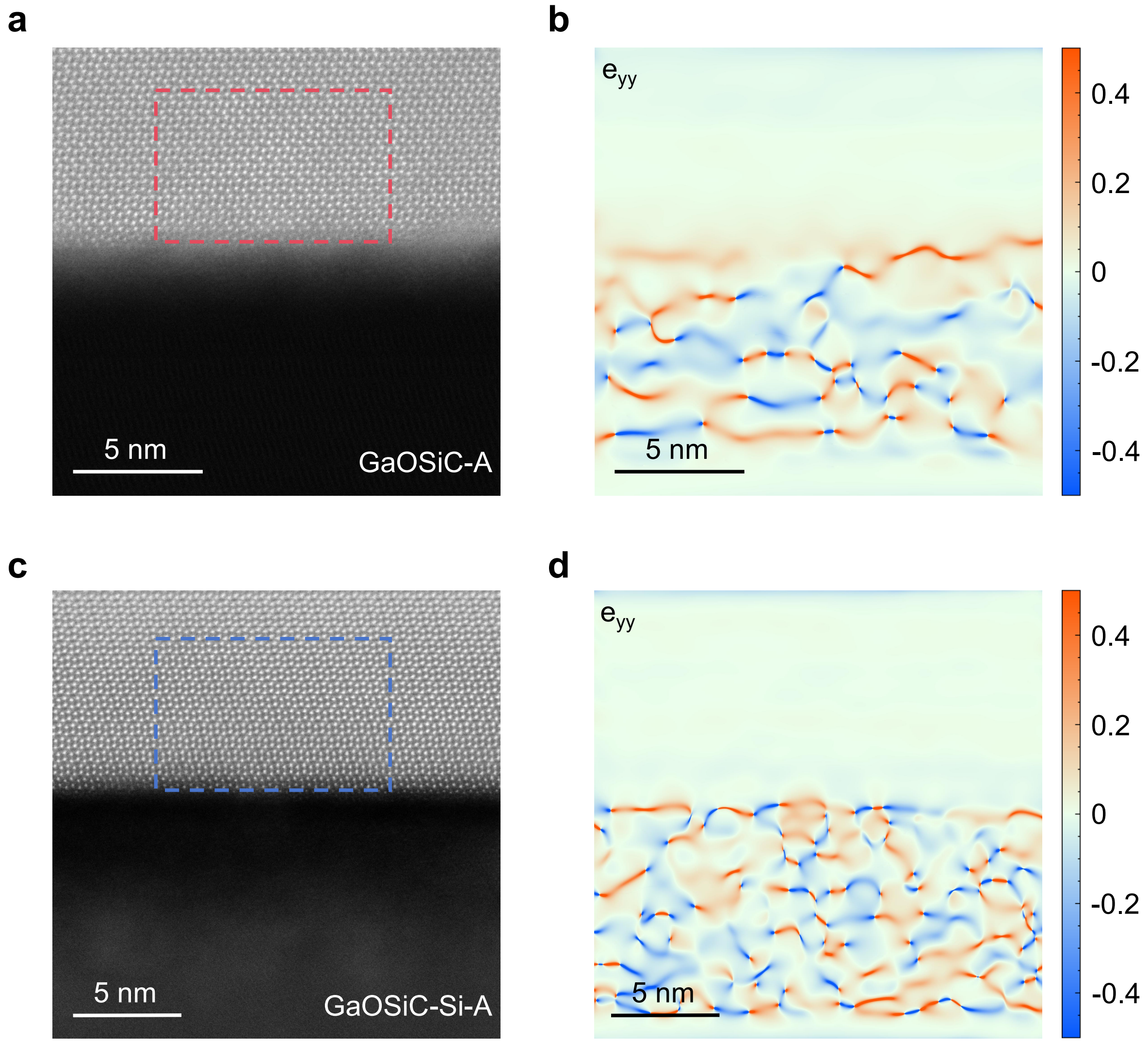

**Figure S5.** Interfacial strain analysis of the annealed samples. a), b) HAADF-STEM image a) and corresponding GPA $\varepsilon_{yy}$ strain map b) of GaOSiC-A. c), d) Corresponding HAADF-STEM image c) and $\varepsilon_{yy}$ strain map d) of GaOSiC-Si-A. The dashed boxes in a) and c) indicate the regions used for atomic-position fitting and plane-resolved strain analysis in Figure 2f.

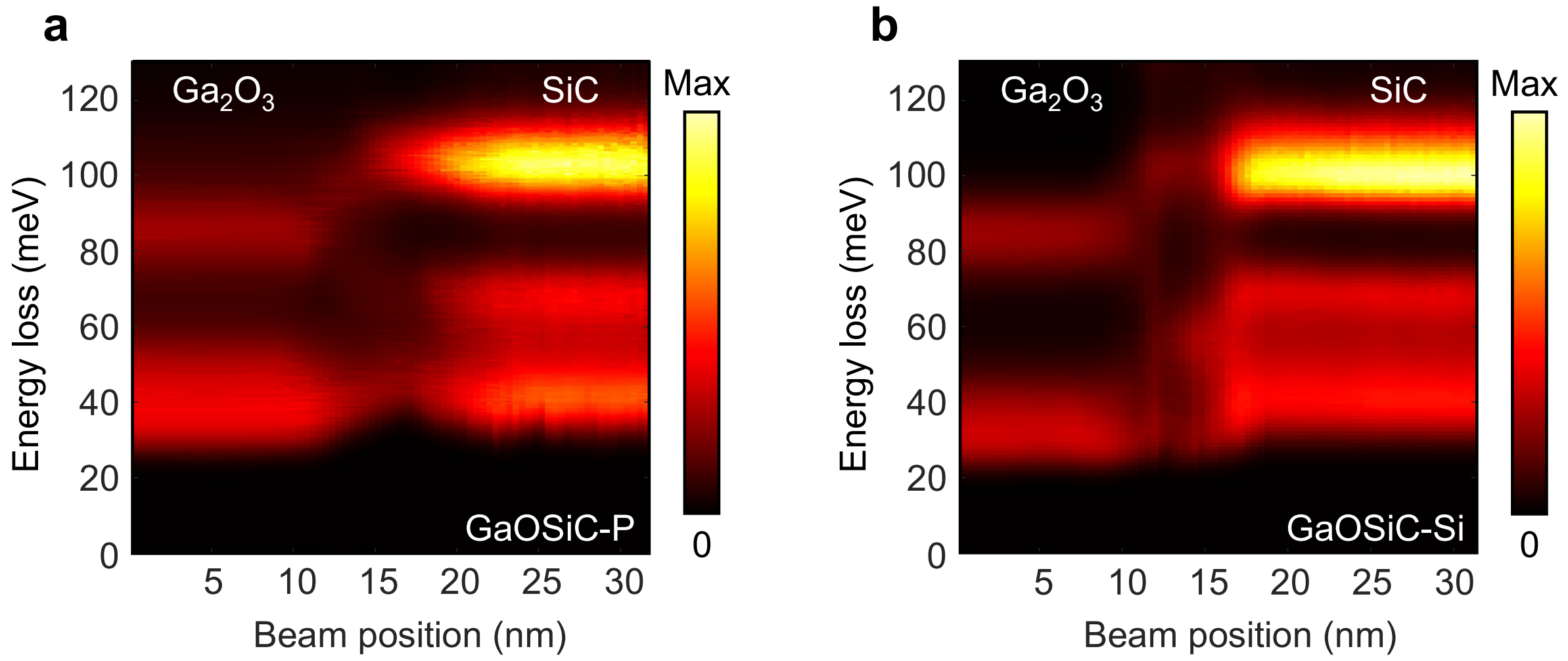

**Figure S6.** Mapping of phonon EELS spectra of the unannealed samples. a), b) Vibrational EELS maps acquired across the GaOSiC-P a) and GaOSiC-Si b) interfaces.

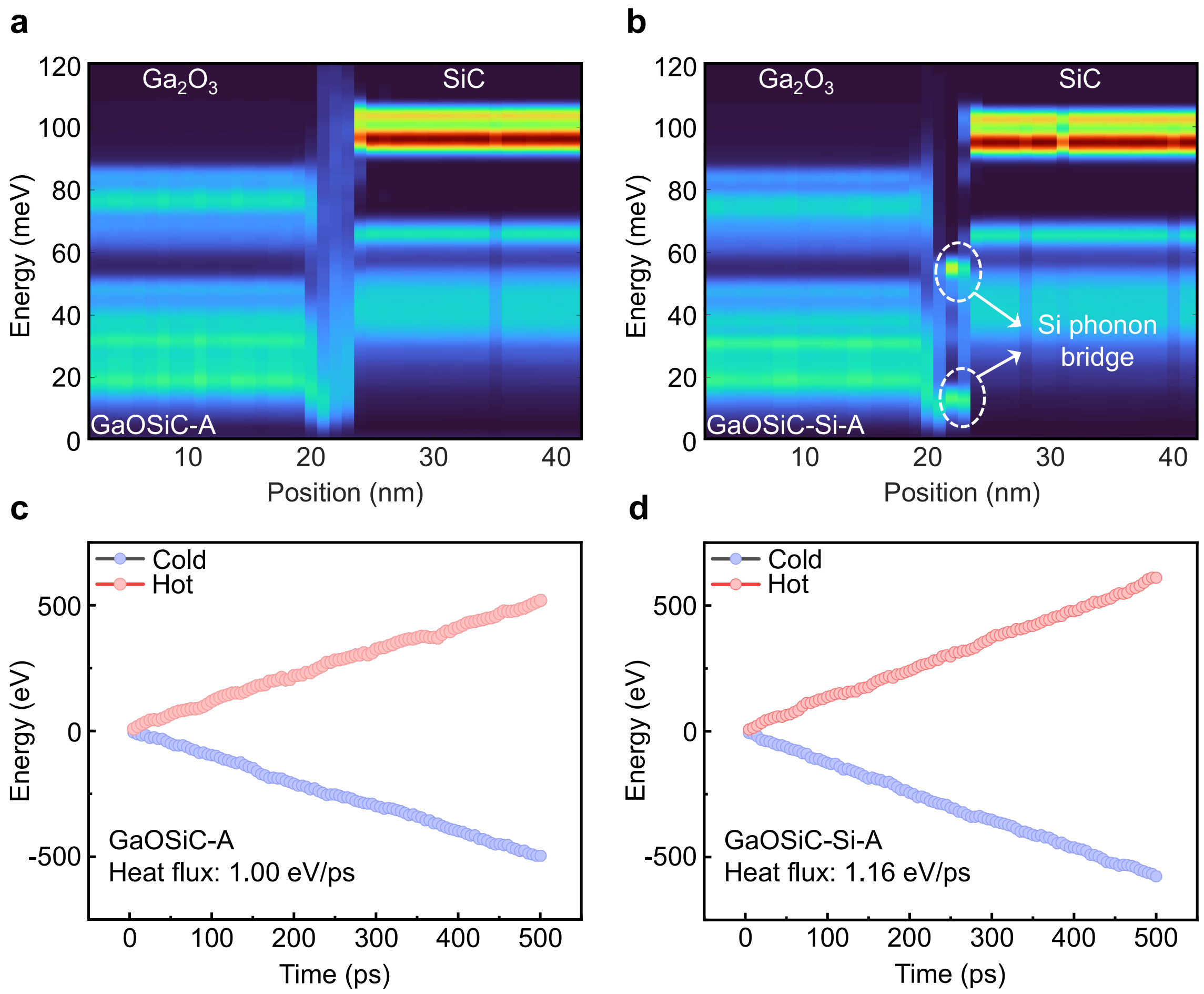

**Figure S7.** MD simulation analysis of interfacial phonon transport. a), b) MD-simulated phonon spectral maps of GaOSiC-A a) and GaOSiC-Si-A b). The circled features in b) indicate calculated interfacial Si vibrational modes. c), d) Calculated accumulative energy of the heat source and heat sink of GaOSiC-A c) and GaOSiC-Si-A d). Linear fitting yields energy-transfer rates of 1.00 and 1.16 eV $ps^{-1}$, respectively.